\documentclass[multphys,vecphys]{svmult}

\usepackage{makeidx}         
\usepackage{graphicx}        
\usepackage{multicol}        
\usepackage[bottom]{footmisc}

\makeindex             

\begin{document}

\title*{UCDs -- a mixed bag of objects}
\author{Michael Hilker\inst{1}}
\institute{Argelander-Institut f\"ur Astronomie, Universit\"at Bonn, Auf 
dem H\"ugel 71, 53121 Bonn, Germany
\texttt{mhilker@astro.uni-bonn.de}}
\maketitle

\section{Introduction}

The realm of dwarf galaxies was extensively studied only in the last three 
decades. Dwarf spheroidals are considered to be the faintest galaxies, 
having baryonic masses comparable to those of bright globular 
clusters ($\sim 10^5 M_{\odot}$), but being 50-200 times more extended.
Thus far one thought that dwarf galaxies were diffuse structures, with the
exception of the compact elliptical M32 which is $\sim$8-10 times smaller than
dwarf ellipticals of comparable luminosities but about 150 times more
luminous than the brightest globular clusters of the Local Group.
This gap in luminosity of compact stellar systems started to be filled in
observationally during the last decade thanks to several large spectroscopic 
surveys in nearby galaxy clusters.
The question arises what are the physical parameters and what is the origin 
of objects in the transition region between dwarf galaxies and star clusters?
What is the smallest compact galaxy? What is the largest globular cluster?
How can a massive {\it cluster} be descerned from a low-mass compact {\it 
galaxy}?

\section{The discovery of compact objects}
 
\subsection{Celebrating the 10-years anniversary of ``UCDs''}

The discovery history of very massive compact star clusters started about
10 years ago. In a small spectroscopic survey of the globular cluster system
of NGC 1399, the central galaxy of the Fornax cluster, Minniti et al. 
\cite{minn98} confirmed a bright compact object as radial velocity member of 
the cluster: {\it `... Note that the objects at $V=18.5$, $V-I=1.48$ (our 
reddest ``globular cluster''), which has $M_V = -12.5$, was identified as a 
compact dwarf galaxy on the images after light-profile analysis (M. Hilker, 
1996, private communication {\rm \cite{hilk98}}) ...''}. 
In another spectroscopic survey, designed as a follow-up of a photometric 
investigation of the surface brightness-magnitude relation of dwarf 
ellipticals in the Fornax cluster, Hilker et al. \cite{hilk99} confirmed in 
1999 two bright compact objects (including the one mentioned before) as 
Fornax members. They proposed that they {\it `... can be explained by a very 
bright GC as well as by a compact elliptical like M32. Another explanation 
might be that these objects represent the nuclei of dissolved dE,Ns ...'}. 
Further they suggested that {\it `... It would be interesting to investigate,
whether there are more objects of this kind hidden among the high surface
brightness objects in the central Fornax cluster ...'}.

Indeed, only one year later, in 2000, a systematic all-object spectroscopic 
survey within in a 2-degree field centred on the Fornax cluster revealed five 
compact Fornax members in the magnitude range $-13.5<M_V<-12.0$ \cite{drin00}
which later, in 2001, were dubbed ``Ultracompact Dwarf Galaxies'' (UCDs) by 
Phillipps et al. \cite{phil01}. Their physical properties were presented in 
a {\it Nature} article by Drinkwater et al. in 2003 \cite{drin03}.

\subsection{A word on nomenclature}

Before the term ``ultracompact dwarf galaxy'' (UCD) was invented, the new 
type of objects was circumscribed in different ways: e.g. compact stellar 
object, compact object, compact stellar system, (extremely) compact dwarf 
galaxy, super-massive star cluster, extremely large star cluster, etc.

After its introduction, the denomination ``ultracompact dwarf galaxy'' became
widely used, but also provoked severe critisism, since it suggests
that these objects have a galaxian origin. 
The term ``ultra-diffuse star cluster'' was opposed by Kissler-Patig 
\cite{kiss04} to demonstrate their link to massive star cluster formation. 
In an attempt to find a neutral description Ha\c{s}egan et al. \cite{hase05} 
named newly discovered compact objects in the Virgo cluster ``dwarf-globular 
transition objects'' (DGTOs).

One should note that star clusters of similar luminosities and sizes as 
UCDs/DGTOs are known in present-day galaxies, like the nuclear (star) 
clusters (NCs) of late-type spirals and dwarf ellipticals and the super-star 
clusters (SSCs) or young massive clusters (YMCs) of merger/starburst galaxies.

In this contribution, I will stay for simplicity with the term UCD to 
describe objects with properties as summarized in the next section. Note
however that I don't suggest that they are of galaxian origin. In contrary,
I will argue that they may be an inhomogeneous group of objects perhaps with
different origins.

\subsection{UCD properties}

Once the existence of UCDs in the Fornax and Virgo clusters was proven
by radial velocity measurements, e.g. 
\cite{hilk99},\cite{drin00},\cite{hase05},\cite{jone06},
their physical parameters became the subject of much study. In particular, 
their sizes, shapes, metallicities, ages, internal kinematics, masses and 
mass-to-light ratios are of interest.
Recent and ongoing observing programmes, employing high-resolution imaging
(HST) and spectroscopy (VLT and Keck) as well as high signal-to-noise 
spectroscopy and deep imaging to faint surface brightnesses, revealed most of 
those parameters for Fornax and Virgo UCDs 
\cite{drin03},\cite{hase05},\cite{rich05},\cite{depr05},\cite{mies06}.
The general properties are listed in Table 1.

\begin{table}[t]
\centering
\caption{General properties of UCDs (LSB = low surface brightness).}
\label{tab1}     
\begin{tabular}{ll}
\hline\noalign{\smallskip}
Luminosity: & $-13.5<M_V<-11.5$ mag \\
\noalign{\smallskip}
Colour: & Fornax: mainly red; Virgo: mainly blue \\
\noalign{\smallskip}
LSB features: & some have LSB envelopes with $80<R_{\rm eff}<120$ pc \\
\noalign{\smallskip}
Shape: & best represented by King+S\'ersic or Nuker profile \\
\noalign{\smallskip}
Effective radius: & $10<R_{\rm eff}<30$ pc \\
\noalign{\smallskip}
Velocity dispersion: & $25<\sigma_0<45$ km~s$^{-1}$ (central value)\\
\noalign{\smallskip}
Mass: & $M=$ 2-9$\times 10^7 M_\odot$ \\
\noalign{\smallskip}
Mass-to-light ratio: \,\, & $M/L_V=$ 2-5; Virgo DGTOs: 2-9 (global value) \\
\noalign{\smallskip}
Metallicity: & Fornax: $-0.5$ dex; Virgo: $-1.2$ dex, [$\alpha$/Fe]$=0.3$ \\
\noalign{\smallskip}
Age: & old: $>$8-10 Gyr \\
\noalign{\smallskip}\hline
\end{tabular}
\end{table}

\section{``The UCD Rush''}

After the first discovery of UCDs in the Fornax cluster, many surveys 
followed to search for UCDs in different environments and towards
fainter magnitudes. 

In the Fornax cluster, Mieske et al. \cite{mies04} identified compact objects
in the brightness range $-12.0<M_V<-10.0$ mag. They found that their 
distribution over luminosity is consistent with an extrapolation of the GC 
luminosity function. Jones at al. \cite{jone06} discovered a sixth bright UCD 
in the central two degrees of the cluster, but could not find any UCD brighter 
than $M_V=-12.0$ in an all-object spectroscopic survey of two adjacent 
2-degree fields. The spatial distribution of GCs and UCDs in Fornax shows that 
UCDs brighter than $M_V\simeq-11.5$ do not seem to belong to certain galaxies,
but rather live in the intra-cluster space of the core region of the cluster.

In the Virgo cluster, Ha\c{s}egan et al. \cite{hase05} identified close to
the central galaxy M87 six DGTOs and 13 DGTO candidates in the magnitude 
range $-11.8<M_V<-10.8$. Three of the DGTOs have global $M/L_V$ of the order
6-9 that cannot be explained by stellar population models. 
Jones at al. \cite{jone06} discovered 9 UCDs with $-13.7<M_V<-11.5$ in a 
2-degree field around M87, widely distributed throughout intra-cluster space. 

In other surveys, nearby groups and distant massive clusters were searched for
UCDs. Around Centaurus A several bright GCs with $M_V>-11.2$ were identified,
e.g. \cite{mart04}, but no massive UCDs. Those seem also to be absent in
other nearby groups. In the very massive lensing cluster Abell 1689, two
twins of M32 were discovered and several massive UCD candidates \cite{mies05}.

\begin{figure}
\centering
\includegraphics[height=9.8cm]{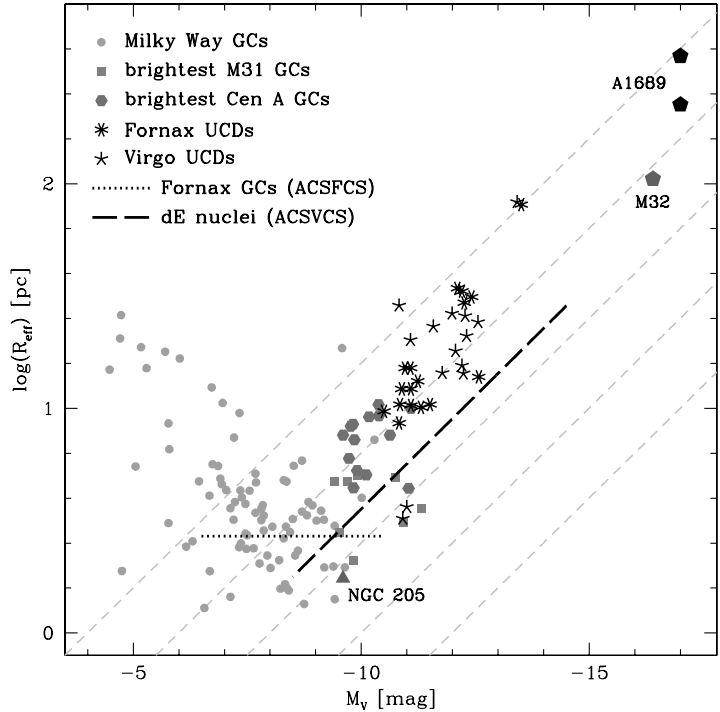}
\caption{Luminosity-size relation for GCs, UCDs, compact ellipticals and 
dwarf ellipical nuclei as indicated in the plot (data taken from the 
literature). 
The diagonal
dashed lines are lines of constant surface intensity. The horizontal dotted
line shows the average effective radius of ``normal'' GCs in the Fornax
cluster, $\langle r_{\rm eff}\rangle = 2.7$ pc.}
\end{figure}

\section{Filling the parameter space of hot stellar systems}

The physical parameters of UCDs have been compared to those of bright star 
clusters (young and old) and galactic nuclei by many authors. It is of special
interest whether UCDs lie on known scaling relations in the parameter space of 
hot stellar systems. The most commonly used parameters for comparison are the
absolute magnitude or mass (if available), the central or effective surface 
brightness, the effective (= half-light) radius, the central velocity 
dispersion, or combinations of such parameters (e.g. $\kappa$-space).

\begin{figure}
\centering
\includegraphics[height=9.8cm]{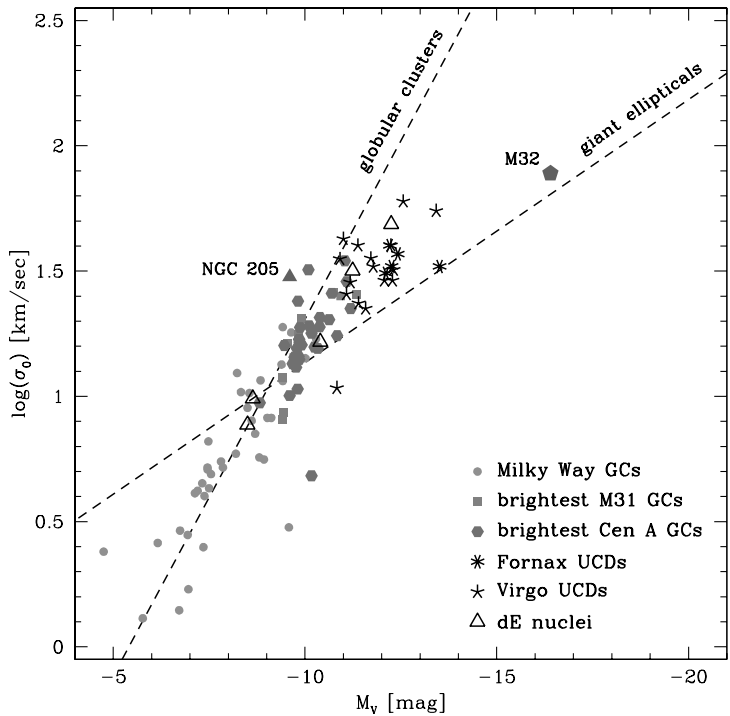}
\caption{Luminosity versus central velocity dispersion for GCs, UCDs, M32 
and dwarf ellipical nuclei as indicated in the plot. The dashed lines 
show the known relations for Galactic globular clusters and giant ellipticals 
(Faber-Jackson relation).}
\end{figure}

Here, the luminosity-size and luminosity-velocity dispersion plane, as well 
as a colour magnitude diagram of Fornax and Virgo UCDs are presented.

\begin{figure}
\centering
\includegraphics[height=9.8cm]{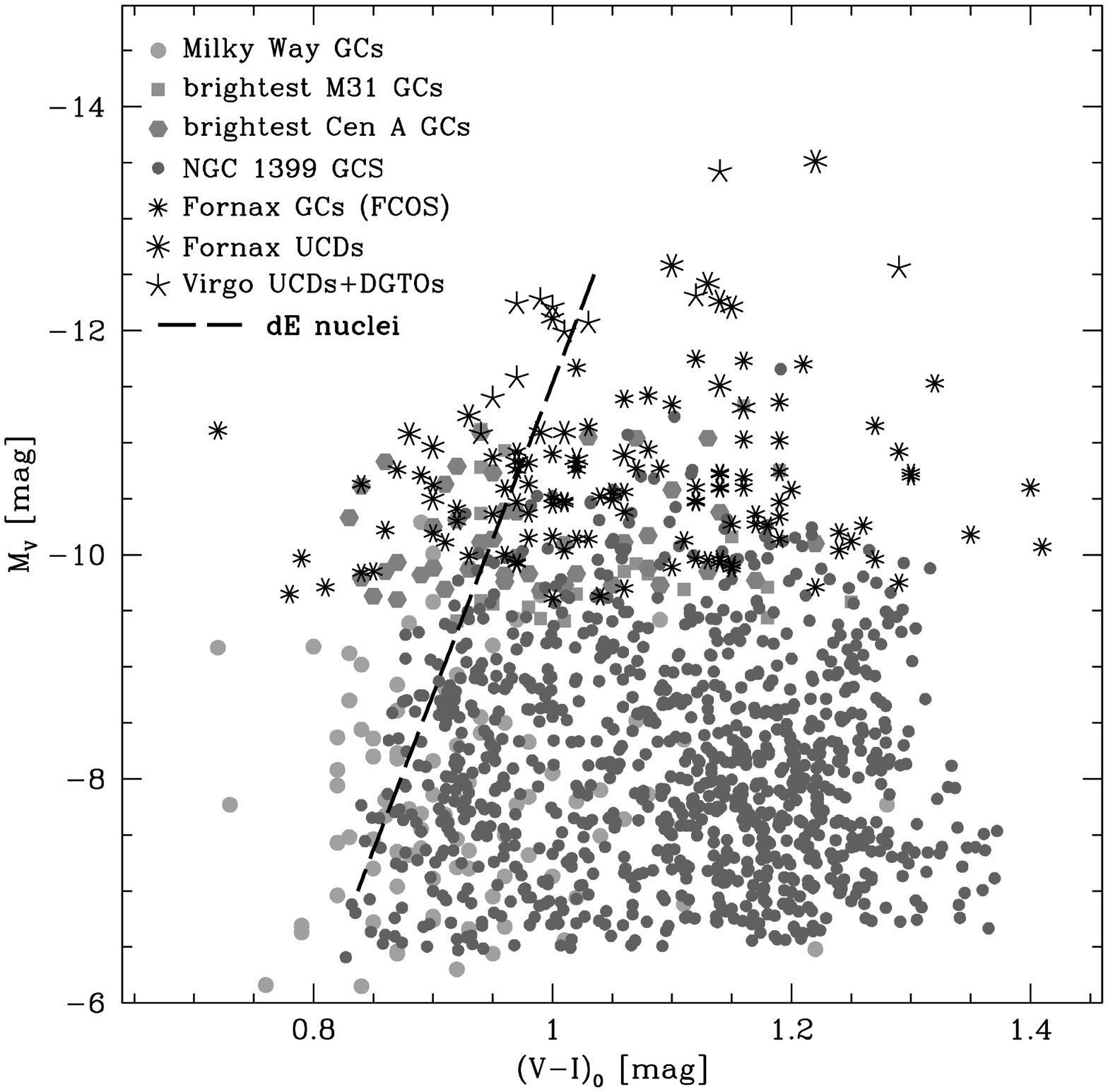}
\caption{Colour magnitude diagram for GCs and UCDs as indicated in the plot.
The dashed line shows the colour-magnitude relation of dwarf elliptical nuclei.}
\end{figure}

Fig.\,1 shows that Galactic GCs with $M_V<-7.5$ and GCs in the
Fornax cluster (dotted line) scatter around a luminosity-independent size 
of about 2.7 parsec. 
Objects brighter than
$M_V\simeq-10.5$, however, follow a luminosity-size relation, approximately
along a line of constant surface density. M32-type galaxies lie on the 
extension of this relation. Also nuclei of early-type galaxies exhibit a 
luminosity-size relation, shifted towards smaller sizes at a given
luminosity. 
Note that young massive star clusters in starburst/merger galaxies follow 
a mass-size relation that is consistent with that of UCDs \cite{kiss06}.

Concerning their internal kinematical properties, compact objects brighter
than $M_V\simeq-10.5$ also seem to deviate from the well defined relation of
``normal'' GCs (see Fig.\,2). In the luminosity-velocity dispersion diagram,
UCDs seem to bend over towards the Faber-Jackson relation of early-type
galaxies.

The colour magnitude diagram in Fig.\,3 exhibits some interesting differences
between Fornax and Virgo UCDs. In the magnitude range $-13.0<M_V<-11.5$, the
Fornax UCDs are dominantly red ([Fe/H]$\simeq-0.5$), whereas the colour of
most Virgo UCDs is blue ([Fe/H]$\simeq-1.2$) and consistent with that of dE
nuclei. Note that the two brightest UCDs are at least twice as luminous as the
second brightest UCD in their respective clusters. Both are metal-rich and 
possess a small envelope of low surface brightness. 

\section{Formation scenarios for UCDs}

Various formation scenarios have been suggested to explain the origin of UCDs.
The three most promising and their implications are:

\vskip1mm

\noindent
{\bf 1)} UCDs are the remnant nuclei of galaxies that have been significantly 
stripped in the cluster environment, also referred to as the ``threshing 
scenario'', e.g. \cite{bass94},\cite{bekk03}. Numerical simulations have 
shown that nucleated dEs can be disrupted in a galaxy cluster potential under
specific conditions and that the remnant nuclei resemble UCDs in their 
structural parameters \cite{bekk03}. Good candidates for isolated 
nuclei are the Galactic globular clusters $\omega$ Centauri \cite{hilk00} and 
M54 as the nucleus of the Sagittarius dSph.

In Fornax and Virgo, the present-day nuclei of dwarf ellipticals are less
massive and more compact than the UCDs. This implies that the progenitor
galaxies must have been very massive dE,Ns or maybe late-type spirals.
The small number of UCDs in both clusters points to a rather selective 
``threshing'' process. The high metallicity of the Fornax UCDs seems to
disfavour this scenario for their origin.

\vskip1mm

\noindent
{\bf 2)} UCDs have formed from the agglomeration of many young, massive star
clusters that were created during merger events, e.g. 
\cite{krou98},\cite{fell02},\cite{kiss06}. Inspired by the massive star
forming knots in the Antennae galaxies, Kroupa \cite{krou98} first showed
that the individual star clusters in complexes can merge to form a large
massive, single star cluster. This work thus
constitutes a prediction of UCD-type objects made right at the time when 
these were discovered. An evolved 
example of such a merged star cluster complex might be the 300 Myr old, 
super-star cluster W3 in NGC 7252 \cite{mara04},\cite{fell05}. 

If the old UCDs in Fornax and Virgo formed like this, the galaxy mergers
must have happened early in the galaxy cluster formation history when the 
merging galaxies were still gas-rich. However, in the case of Fornax, these
early mergers must have already possessed close to solar metallicity gas.
The small number of UCDs would imply that only the most massive star cluster
complexes survived as bound systems.

\vskip1mm

\noindent
{\bf 3)} UCDs are the brightest globular clusters and were formed in the same
GC formation event as their less massive counterparts, e.g. 
\cite{mies04},\cite{mart04}. The smooth shape of the bright end of the
GC luminosity function (no excess objects!) might support this scenario.

The most massive GCs then supposedly formed from the most massive molecular
clouds (MCs) of their host galaxy, where more massive galaxies were able to
form higher mass MCs (as e.g. M87) than lower mass galaxies (as e.g. M31). 
The luminosity-size relation of the most massive clusters suggests that there
is a break of the formation/collapse physics at a critical MC mass. 
In Fornax, the formation of the most massive GCs was connected
to that of the metal-rich bulge GCs, whereas in Virgo they must have formed
with the metal-poor GCs, if this scenario would be correct.

%
%

\vskip1mm

In an attempt to unify the ideas of the different formation scenarios one might 
think of the following generalized scheme of massive star cluster formation:
GCs with a mass of $\le5\times 10^6 M_\odot$ are ``single-collapse'' GCs 
(SCGCs), i.e.  their stars share a single age and metallicity. At a critical
mass of $\simeq 5\times 10^6 M_\odot$ the formation physics changes 
to ``multiple-collapse'' GCs (MCGCs) because the giant MC fragments into 
massive clusters as it contracts.
The growth of SCGCs to MCGCs through merging can happen on different time
scales. An immediate growth ($\sim 10$-100 Myr) would correspond to the 
situation in super-star cluster complexes in mergers. The stars formed in the 
resulting MCGC then would have the same age and probably similar metallicities,
although MCGCs would also be able to capture a substantial amount of older 
field stars from the host galaxy \cite{krou98}.
A successive growth over gigayears reflects the situation in nuclear star
clusters. Nuclei of dwarf ellipticals could have formed via the merging of
GCs which not necessarily had all the same age and/or metallcity. Another way
of growing a nucelar star cluster is via episodic star formation of infalling
gas in the centre of a gas-rich galaxy. The stars of those MCGCs are supposed
to show a spread in ages and metallicities.
Finally, through whichever channel the MCGCs formed, their evolution in
the tidal field of a dense galaxy cluster over a Hubble time could have given
rise to the population of old, isolated, massive, compact stellar systems
we nowadays call UCDs. 

\section{Summary and Outlook}

The name ``ultracompact dwarf galaxies'' (UCDs) collects/paraphrases old
stellar systems in the transition region of globular clusters and compact
dwarf galaxies ($-13.5< M_V < -11.5$, 2-9$\times10^7 M_\odot$, $10<r_h<30$ pc,
$25<\sigma_0<45$ km~s$^{-1}$). The known UCDs are found in the cores of galaxy 
clusters and are not concentrated towards individual galaxies unlike most of 
the GCs.

While we have good ideas on their possible origin, there are many questions
left to answer concerning the nature of UCDs. Some 
important ones are: Do UCDs have multiple stellar populations? Why do some
UCDs have high M/L ratios? Is this due to tides? Or do they contain dark 
matter? Is there tidal structure around UCDs? Do UCDs harbour black holes? 
What are the kinematics of UCDs within their host clusters?

Some of these questions will be answered in the next years with the help of
ongoing and future observing programmes. The results will bring more light
into the nature of these enigmatic objects.



\printindex
\end{document}